\documentclass[12pt]{article}
 \usepackage{latexsym,epsfig,color,times}
\usepackage{colordvi}
\usepackage{amsmath}
\usepackage{latexsym,graphicx,color,times}
\usepackage[normalem]{ulem}
\usepackage[portrait,margin=1in]{geometry}

\newcommand{\be}{\begin{equation}}
\newcommand{\ee}{\end{equation}}

\def\Rdd{\Red}
\def\Rdd{\Black}
\def\Cyn{\Cyan}
\def\Cyn{\Black}

\def\hline{\centerline{\Red{\rule{6in}{1.0pt}}}}

\def\Alf{Alfv\'en }
\def\rfig#1{Fig.\ref{fig:#1}}

\def\req#1{(\ref{eq:#1})}

\def\macname{hankrs2}
\def\macname{hank_strauss}
\def\fignf12{./pix}

\def\figdircc1{/Users/{\macname}/Documents/progs/m3dc1/plots}

\def\figdir{./}

\begin{document}

\begin{center}
{\Large{\bf Models of Tokamak Disruptions}} \\
{\large\bf H. R. Strauss}
\end{center}
\begin{center}
 {HRS Fusion} \\
 {West Orange, NJ 07052 USA}\\
 {hank@hrsfusion.com} \\
\end{center}

\abstract{ \it
Disruptions are a  serious issue in tokamaks. 
In a disruption, the thermal energy is lost by means of an instability which
could be a resistive wall tearing mode (RWTM).
During  precursors to a disruption,
the plasma edge region cools, causing the current to contract.
Model sequences of contracted current equilibria are given, and their stability
is calculated. 
A linear stability study shows that there is a maximum value of edge $q_a \approx 3$ for
RWTMs to occur. This also implies a minimum rational surface radius normalized to plasma
radius from RWTMs to be unstable. 
Nonlinear simulations are performed using a
 similar model sequence derived from an equilibrium reconstruction.  There is a striking
difference in the results, depending on whether the wall is  ideal or resistive. 
With an ideal wall, the perturbations saturate at moderate amplitude, causing a minor
disruption 
without a thermal quench. With a resistive wall, there is a major disruption with a  thermal quench, if
the edge $q_a \le  3.$ There is a sharp transition in nonlinear behavior at $q_a = 3.$ 
This is consistent with the linear model and
with experiments. 
If disruptions are caused by RWTMs,
then devices with highly conducting walls, such as the International
Tokamak Experimental Reactor (ITER) will experience much milder,
tolerable, disruptions than presently predicted.
}

\vspace{1cm}

\section{Introduction}

Disruptions are a  serious issue in tokamaks, the leading magnetic  fusion device.
Recent work has identified disruptions in JET \cite{jet21}, ITER \cite{iter21},
DIII-D \cite{d3d22}, and MST \cite{hurst,mst23} as possibly caused by 
 resistive wall tearing modes (RWTMs)
\cite{finn95,gimblett,bondeson}.
It was shown in numerical simulations that RWTMs are able to cause a complete thermal quench (TQ).
An onset condition for a RWTM is that with an ideal wall, it would be a stable tearing mode (TM).
This is consistent with 
simulations and experimental data.

An object of this paper is to show that experimental  conditions for 
 tokamak 
disruptions are also conditions for RWTM instability.  
Models are presented which shows that RWTMs can occur when edge
cooling causes contraction of the current. 
The model shows that the rational surface radius $r_s$ must be sufficiently
close to the resistive wall, or that the edge $q_a $ be sufficiently low for RWTMs to occur.
The $q_a$ value for RWTM instability is  $q_a \stackrel{<}{\sim}  3.$  
This is consistent with experiments, which 
 generally run with $q_a > 3$ to avoid disruptions \cite{iter1999}.
The onset condition is studied linearly using a sequence of equilibrium models
with varying $q_a$ and current contraction,
which can be studied semi analytically.
The onset condition  is also studied nonlinearly, with a more realistic sequence of equilibria
with varying $q_a$. The equilibria are constructed from an MST equilibrium 
reconstruction \cite{mst23}. 
The marginal stability condition for RWTMs is consistent between
the two models. The nonlinear simulations illustrate the basic result that RWTMs can
cause a thermal quench. There is a striking difference in the simulations with ideal 
and resistive walls. Nonlinear simulations of the same equilibria
with an ideal wall boundary condition only obtain a minor disruption without a  TQ. 
Simulations with a resistive wall obtain a major disruption and TQ, when $q_a \le 3.$ For a resistive
wall, there is a sharp transition in the simulations at the RWTM onset condition
$q_a = 3.$ For $q_a > 3,$ the behavior is similar to having an ideal wall. 

Disruptions are generally preceded by precursors, 
which typically involve tearing modes.
Numerous causes of precursors in JET
have been identified \cite{devries11}, which lead to locked modes.
These  include
 neoclassical tearing modes (NTM) \cite{lahaye},
and  radiative cooling by impurities \cite{pucella}.
Nearly all JET disruptions are preceded by locked modes,
 but they are not the instability causing the
thermal quench.  Rather,
the locked mode indicates an ``unhealthy" plasma which may disrupt
\cite{gerasimov2020}.
Locked modes are also disruption precursors in DIII-D \cite{sweeney2017,sweeney}.
The locked modes are tearing modes. They can overlap
and cause stochastic thermal transport in the plasma edge region.

During the locked mode phase, 
edge transport and cooling modify the edge temperature and current.
The drop in the edge temperature
causes the current to contract, while the total current stays 
constant. The result has been called \cite{schuller}  a
``deficient edge".  It has also been described \cite{sweeney}  as ``$T_{e,q2}$ collapse", a minor
disruption of the edge.
The edge cooling causes the resistivity in the edge
to increase, which increases the growth rate of TMs and RWTMs.

A condition for disruptions is that the $q = 2$ magnetic surface is sufficiently
near the plasma edge. This is been documented in DIII-D \cite{sweeney2017}.
It was found that disruptions require the $q = 2$ rational surface radius 
$r_s > 0.75 r_a,$ 
where $r_a$ is the plasma minor radius.

\Rdd{The TQ times found in RWTM simulations are consistent with experiment.  
In \cite{iter1999} a  JET  TQ time was given as $0.7 ms,$ while 
$1.5 ms$ was found in simulations \cite{jet21}. This is in reasonable agreement  with the TQ times
calculated \cite{jet21} from experimental shots listed in the JET ITER - like wall 
2011 - 2016 disruption database \cite{gerasimov2020}.
The TQ time in a DIII-D simulation \cite{d3d22} agreed with the experimental data.
The TQ time in an MST simulation \cite{mst23} exceeded the experimental pulse time, consistent with
the lack of disruptions observed.
It was noted \cite{iter1999}  that TQ times scaled linearly with machine size,
 which suggests they are proportional to the  \Alf time. This does not 
rule out a dependence on resistive wall penetration time $\tau_{wall}$, because all the
experiments had similar wall times. 
An exception is MST \cite{mst23}, which has a 
wall time of $800 ms$.} 


\section{Linear model} 

Linear MHD stability is studied using a set of model equilibria in a straight periodic cylinder..
The model equilibria are modified FRS \cite{frs} profiles.
The current density is
\be j(r) = \frac{2}{q_0} (1 + r^{2n})^{-(1 + 1/n)} \label{eq:jfrs} \ee
\Cyn{It is normalized to $B/ R$, where $B$ is the total magnetic field, $2\pi R$ is the
periodicity length,  and radius $r$ normalized to the plasma radius.}
A peaked profile has $n = 1,$
rounded, $n = 2,$ and flattened, $n = 4.$ In this model $n$ is a real
number, not restricted to  an integer.
In order to cut off the current at $r = r_c,$ a constant
$c_r$ is subtracted, with
\be c_r = (1 + r_c^{2n})^{-(1 + 1/n)} \label{eq:cr} \ee
where $r_c$ is the maximum radius of nonzero current normalized to plasma radius..
\be
j(r)   =              \begin{cases}
 (2c_0/q_0)[ (1 + r^{2n})^{-(1 + 1/n)} - c_r] &  r < r_c  \\
  0  &  r \ge r_c.
\end{cases} \label{eq:cases} \ee
The factor  $c_0 = 1 / (1 - c_r)$ keeps $j(0)$ independent of $r_c.$
This gives a $q$ profile
\be
q(r)   =              \begin{cases}
 (q_0/c_0)[ (1 + r^{2n})^{-1/n} - c_r]^{-1}  &  r < r_c  \\
 q(r_c)(r/r_c)^2  &  r \ge r_c.
\end{cases} \label{eq:cases-q} \ee
Note that the total current is given by
\be I 
= r_a^2 / q_a
= r_w^2 / q_w,
 \label{eq:jtot} \ee
where $q = q_a$ at the plasma edge $r_a = 1,$  or by
$q_w$ is the value of $q$  at the wall radius
normalized to plasma radius $r_w.$

Sequences of equilibria during a precursor are modeled by keeping
$q_0 = 1,$ and by fixing $q_a$ to have  constant $I$.
During the sequence, $r_c$ is decreased. This causes the profile parameter  $n$
to increase,
in order to maintain constant $q_0, q_a.$
Current shrinking and broadening occur simultaneously.
The change in linear stability during this model sequence is investigated,
with both ideal and no wall boundary conditions. Resistive wall tearing modes
are tearing stable with an ideal wall, and unstable with no wall.

\Cyn{Rotation is not included. The mode is assumed locked.}

 The ideal wall tearing stability parameter $\Delta'_i$ and
the no wall tearing stability parameter $\Delta'_n$ are calculated in
cylindrical geometry. 
RWTMs have  \cite{jet21,d3d22,mst23}
$\Delta'_i < 0$, and
$\Delta'_{n} > 0.$
Linear magnetic perturbations satisfy \cite{finn95,cheng,frs,fkr}
\be \frac{1}{r}\frac{d}{dr}r\frac{d\psi}{dr} - \frac{m^2}{r^2} \psi  =
\frac{m}{r} \frac{dj}{dr} \frac{m/q - n}{[(m/q - n)^2 + m^2 \delta^2]} \psi
\label{eq:eq} \ee
where the singularity at the rational surface is regularized \cite{cheng},  with
$\delta = 10^{-4}.$
In case $r_c < r_s,$ the right side of \req{eq} vanishes for $r > r_c$, so there is
no singularity at $r_s$  and
$\psi \propto r^{\pm m}.$
Here $(m,n)$ are the poloidal and toroidal mode numbers of
a perturbation $\psi(r)\exp(im\theta - in \phi),$
using a large aspect ratio approximation.
\begin{figure}[h]
\vspace{.5cm}
\centering
 \includegraphics[height=5.0cm]{\figdir/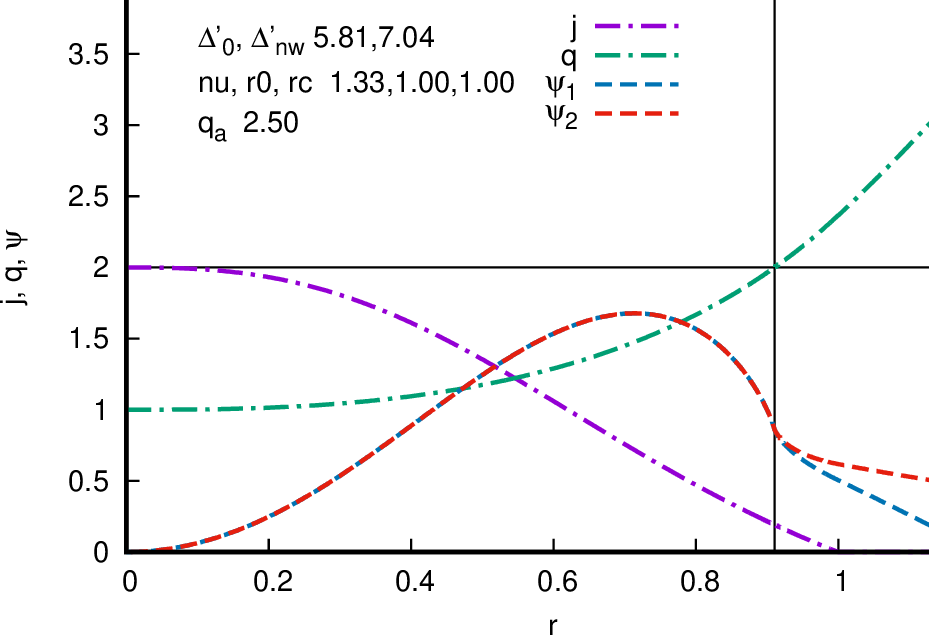}(a)
 \includegraphics[height=5.0cm]{\figdir/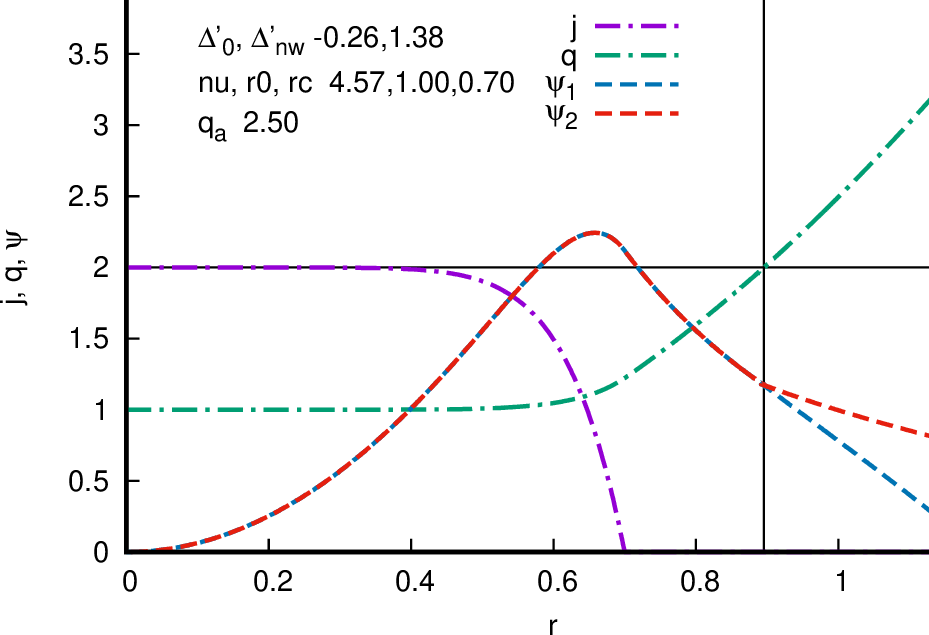}(b)
\caption{\it
$\psi,$ $j$, and $q,$  with $\psi$ for ideal $(\psi_1)$  and no wall
$(\psi_2)$. (a)
tearing mode unstable. The current is nonzero for $r < 1.$ (b) RWTM unstable.
The current
is non zero for $r  < r_c =  .75.$ The current profile is flattened
so the total current is almost the same as in (a). In both
cases $q_0 = 1.$
}
 \label{fig:itlock}
 \end{figure}

Solving with a shooting method, there are two boundary conditions:
integrating outward from $r = 0,$ and inward from $r = r_w,$ the
wall radius.
The boundary conditions at the origin are
$\psi(0) = 0, d\psi/dr(0) = 0,$ since $\psi \sim r^m,$
with $m \ge 2.$
At the wall $r=r_w,$  an ideal wall boundary condition is
$\psi(r_w) = 0,$ $d\psi/dr(r_w) = 1.$
A resistive wall (or no wall)  boundary condition is
$\psi(r_w) = 1,$ $d\psi/dr(r_w) = -(m/r_w) \psi(r_w).$
The value of $\Delta'$ is calculated at $r_s$ at
which $q(r_s) = m/n,$
\be \Delta' = \frac{\psi'_+(r_s)}{\psi_+(r_s)} - \frac{\psi'_-(r_s)} {\psi_-(r_s)}  \label{eq:deltap} \ee
where $\psi' = d\psi/dr$,
$\psi_-$ is the solution integrated outward from $r = 0,$ and
$\psi_+$ is the solution integrated inward from $r = r_w.$
For an ideal wall, denote $\Delta' = k_\perp \Delta_i,$ while for no wall,
$\Delta' = k_\perp \Delta_n,$  where $k_\perp = m /r_s.$
The RWTM instability condition  is $\Delta_i \le 0,$
$\Delta_n \ge 0.$

\Cyn{In the following, $(m,n) = (2,1)$, which are the dominant mode numbers in
typical disruptions.}

The effect of the boundary conditions is illustrated in \rfig{itlock}(a),(b).
The plots show $j(r),$ $q(r),$ and $\psi(r)$ for
both ideal wall $(\psi_1)$ and resistive wall $(\psi_2)$.
The plasma boundary is $r_a = 1,$ and
the wall is at $r_w = 1.2.$
The values of $\psi$ were normalized so that $\psi_+(r_s) = \psi_-(r_s).$
In each figure the two cases have the same profiles of $j$ and $q,$ as well
as the same $\psi_-$. The profiles of $\psi_+$ differ. The no wall
boundary condition produces a more positive value of $\Delta',$
\be \Delta_{n} - \Delta_i = \Delta_x \ge 0. \label{eq:deltax1} \ee
\rfig{itlock}(a),(b) have different $j(r)$ profiles.
Both cases have approximately the same total current $J$
and have $q_0 = 1,  q_a = 2.5.$ 
In \rfig{itlock}(a), $j$ is non zero for $r <  1.$
In \rfig{itlock}(b), $j$ is non zero for $r < r_c = 0.70.$
There is a marked difference in $\Delta'.$
The case
in \rfig{itlock}(a) is unstable to a tearing mode, while the second case
in \rfig{itlock}(b)
is unstable to a RWTM.
This supports the conjecture that suppressing the current in the plasma
edge region destabilizes the RWTM. The RWTM also requires that $r_s$ be
sufficiently close to $r_w,$ so that $\Delta'_i$ can become less than zero.

\begin{figure}[h]
\vspace{.5cm}
\begin{center}
 \includegraphics[width=7.5cm]{\figdir/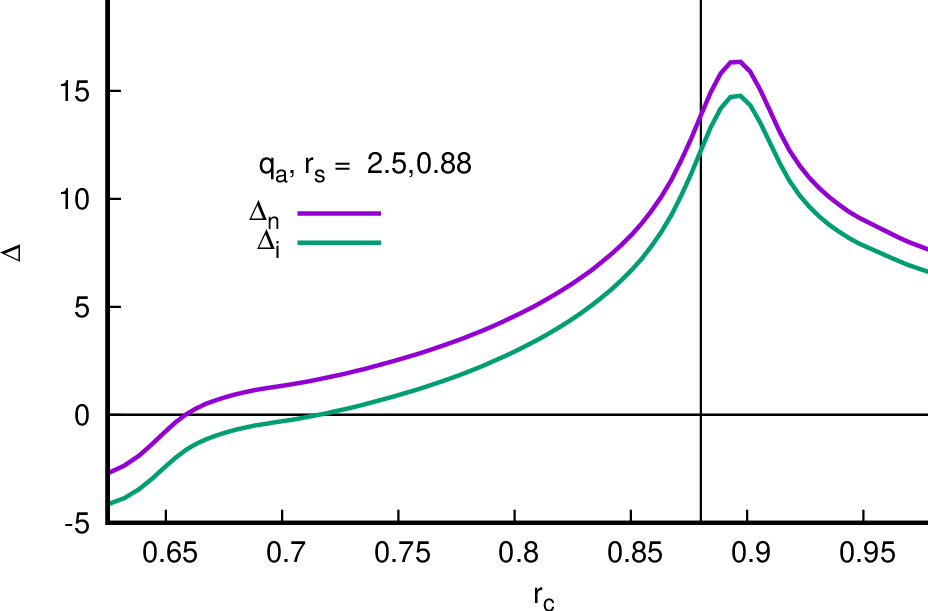}(a)
 \includegraphics[width=7.5cm]{\figdir/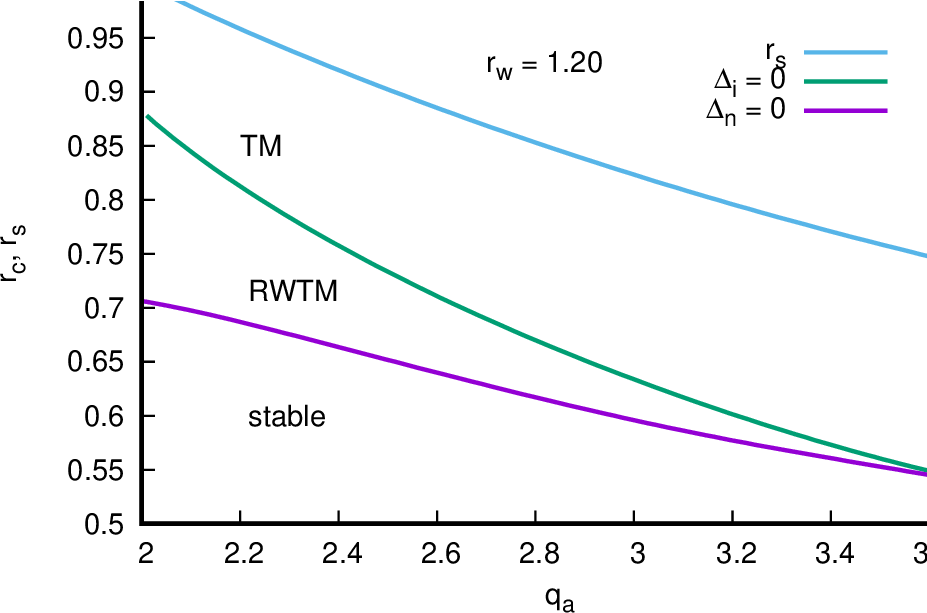}(b)
\end{center}
\caption{\it
 (a)  $\Delta_n, \Delta_i,$ 
as a function of
$r_c,$ for  $q_a = 2.5.$ 
$\Delta_i < 0$ for $r_c < 0.8,$ and $\Delta_n < 0$ for $r_c < 0.7.$
(b) Curves of $\Delta_i(q_a,r_c) = 0$ and $\Delta_n(q_a,r_c) = 0$.
The curves join  at $q_a \approx 3.5,$ $r_s = .75,$ $r_c = 0.57.$ 
}
 \label{fig:qarc}
 \end{figure}

\rfig{qarc}(a) shows how $\Delta_i,\Delta_n$ vary with the current limiting radius
$r_c.$ The rational surface radius $r_s = .88$ and $q_a = 2.5$ are constant.
As $r_c$ decreases, at first the TM is destabilized,
as $\Delta_i,\Delta_n$ increase.
For more contraction and smaller $r_c,$ \Rdd{ the TM is stabilized and the RWTM is
destabilized, as }
the values of $\Delta_i,\Delta_n$ decrease, with
$\Delta_n > \Delta_i.$
For $r_c \le  0.71,$ $\Delta_i \le 0.$ This is the onset condition for a RWTM.
\Rdd{Further contraction stabilizes the RWTM.}
For $r_c \le  0.66,$ $\Delta_n \le 0.$ This implies the RWTM is stabilized.
There is a range of $0.71 \ge r_c \ge 0.66$ in which the RWTM is unstable.

\rfig{qarc}(b)
shows how the marginal $\Delta_i,\Delta_n$ values vary with $q_a.$ 
The critical values of
$r_c$ are found for both $\Delta_i =0, $
and for $\Delta_n =0. $
As in \rfig{qarc}(a) there is a
gap in $r_c$ between RWTM instability and stability.
The curves join  at $q_a \approx 3.5,$ $r_s = .75,$ $r_c = 0.57.$
For $q_a > 3.5,$ 
the RWTM is stable. 
The condition for RWTM instability in the model 
 agrees well  with condition for disruptions  in a 
DIII-D database \cite{sweeney2017}. 
\begin{figure}[h]
\vspace{.5cm}
\begin{center}
 \includegraphics[width=7.5cm]{\figdir/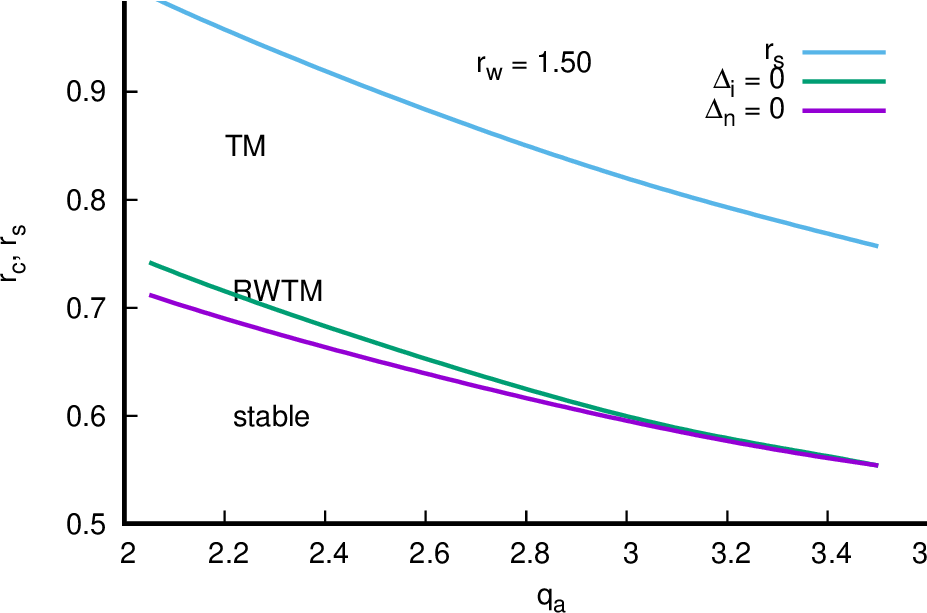}(a)
\includegraphics[width=7.5cm]{\figdir/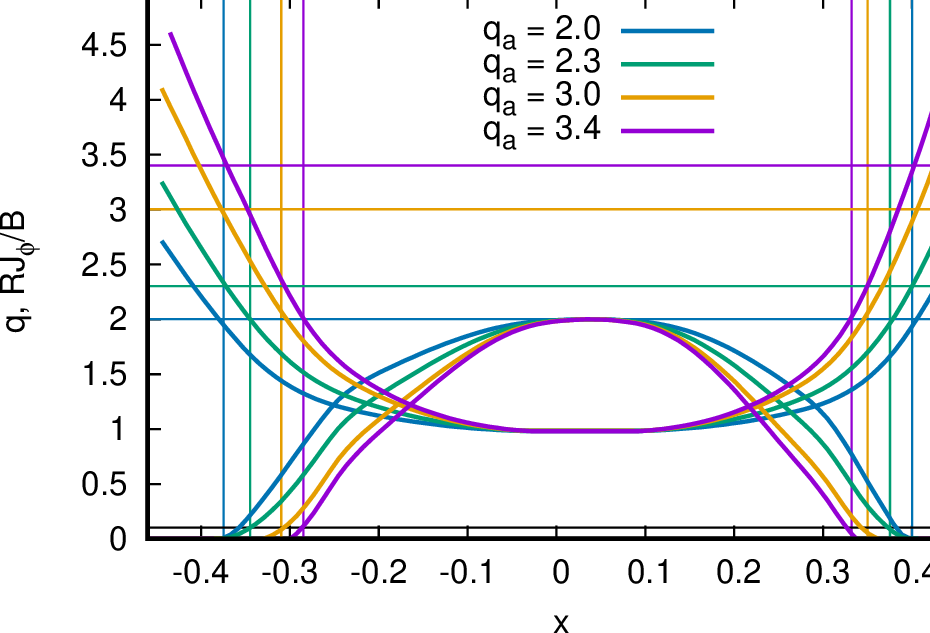}(b)
\end{center}
\vspace{-.5cm}
\caption{\it
(a) Curves of $\Delta_i(q_a,r_c) = 0$ and $\Delta_n(q_a,r_c) = 0$, with $r_w = 1.5.$
There is only a small RWTM unstable region.
(b) $q$ and $RJ_\phi/B$ profiles of a sequence of equilibria derived from
MST,  with $q_a = 2.0, 2.3, 3.0, 3.4.$ At the
respective $(2,1)$ rational surfaces $r_s,$ 
the current density has less than $5\%$ of its value on axis. The current profile is
progressively contracted as $q_a$ increases. 
The minimum $r_s = 0.8 r_a.$
}
\label{fig:qa1}
\end{figure}

\begin{figure} 
\vspace{.5cm}
\begin{center}
\includegraphics[width=5.5cm]{\figdir/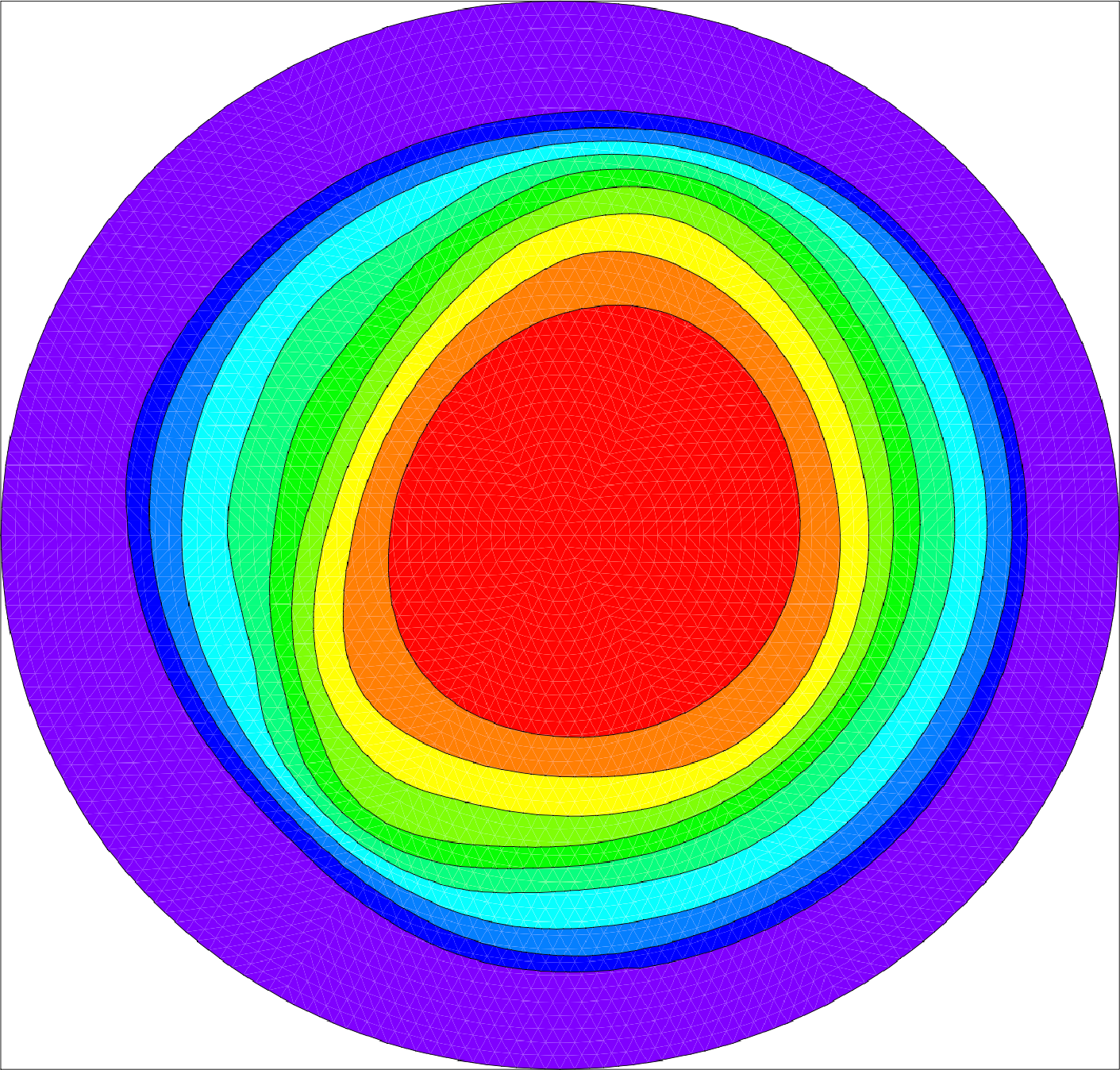}(a)
\hspace{.5cm}
\includegraphics[width=5.5cm]{\figdir/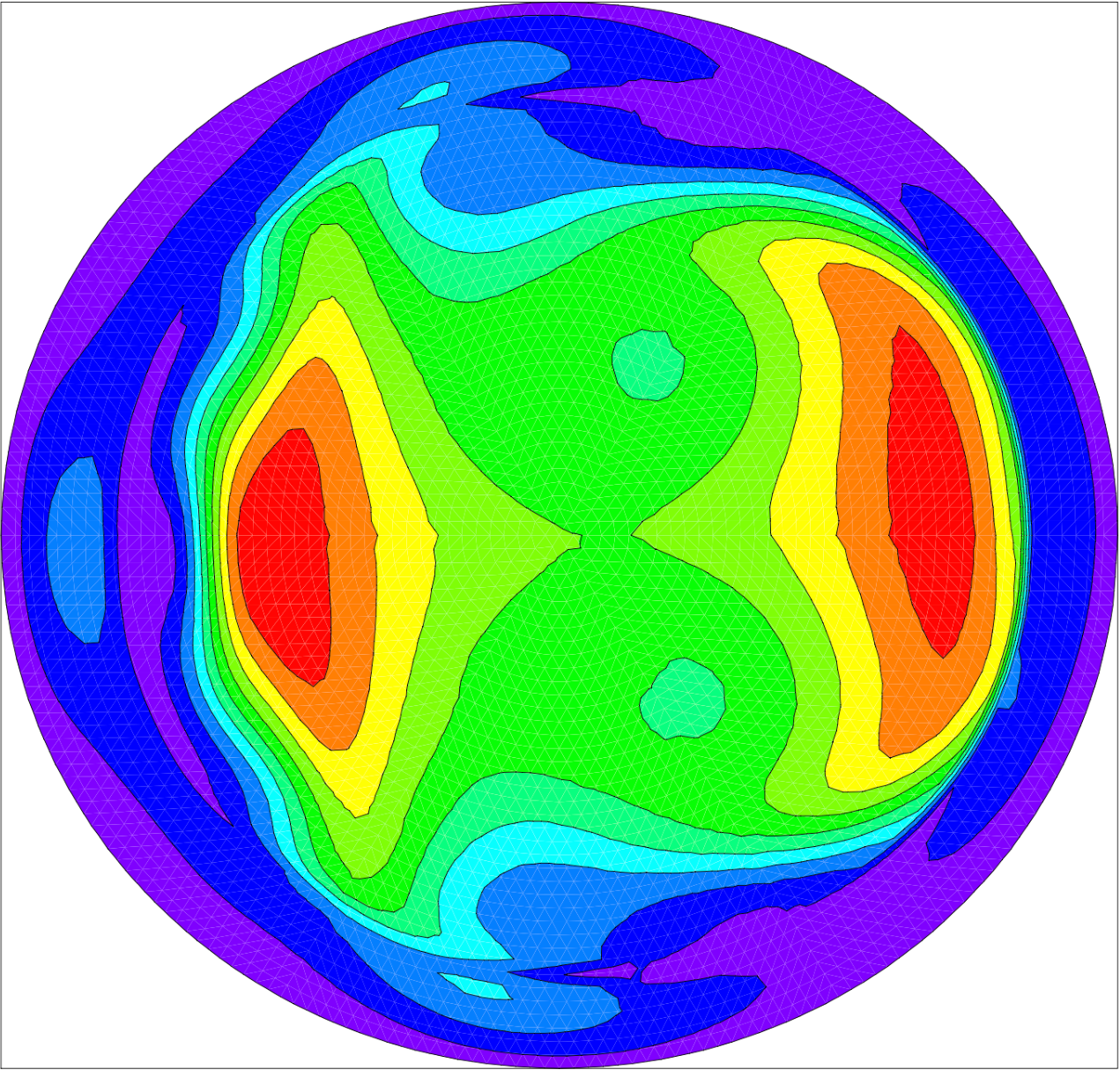}(b)
\end{center}
\caption{\it
(a) Temperature   $T$ contours at 
 $t = 6059 \tau_A,$ for $q_a = 3.0,$ 
with ideal wall, and saturated $(2,1)$, $(3,2)$  modes.
 (b) temperature $T$ at $t = 6390 \tau_A,$
 $q_a = 3.0,$ with 
resistive wall, $S_{wall} = 10^3,$
 and a large predominantly $(2,1)$ island structure.
}
 \label{fig:cont}
 \end{figure}

In DIII-D and the linear examples, 
$r_w / r_a  =  1.2.$
For larger $r_w/r_a,$ the region of RWTM instability in \rfig{qarc}(b) becomes 
smaller. \rfig{qa1}(a) shows the zero curves of $\Delta_i,\Delta_n$ for $r_w = 1.5.$  
There is little  difference in stability between ideal and resistive wall
tearing modes. For larger $r_w/r_a \approx 1.8$, the RWTMs disappear. 
The  modes become no wall tearing modes, whose properties will be
studied elsewhere.

\section{Nonlinear thermal quench} 

Nonlinear simulations were performed with the  M3D resistive 3D MHD code  \cite{m3d} 
including a resistive wall
\cite{pletzer},  using  more realistic
equilibria.
The equilibria were derived from an MST equilibrium reconstruction
\cite{mst23} with $q_a = 2.0,$ which was unstable to a RWTM. The major radius is $R = 1.5m,$
and the wall minor radius is $r_w = 0.48 m$ in the simulations. The 
current was constricted away from the wall, with $r_a = 0.4 m,$ 
to make the equilibrium more unstable.  
A set of equilibria was prepared with varying $q_a$. 

The initial 
 current density was   $C_1 = R \nabla R^{-1} \cdot \nabla \psi_1,$ where
$\psi_1$ is the magnetic flux. In order to vary $q_a,$ 
a new equilibrium was produced with $C_2 = C_1$ initially.
The current $C_2$ was set to zero outside a particular value of normalized radius
$\rho > \rho_a$ which was a  flux surface of $\psi_1.$
In the following, $\rho_a = .83 \rho_w,$ so $r_w / r_a = 1.2.$
The value $C_2(\rho_a)$ was subtracted from $C_2$ for $\rho \le \rho_a.$
The pressure was modified in the same way. The initial 
volume averaged $\beta \approx  5 \times 10^{-3}$, so the presure modification
had a negligible effect.
A new current density $C = c_1 C_2 + c_2 C_2^2$  with constants $c_1,c_2,$
was constrained to have $C(0) = 2 $ on axis, which gives $q_0 \approx 1$ on axis
and volume integral $\int C dV  =  c_3 \int C_1 dV,$ where $c_3$ is a given constant.
This gives two linear equations for $c_1,c_2.$
The constant $c_3$ was chosen to
give a range of $q_a$ values, where $q = q_a$ is the value at the plasma edge
$\rho = \rho_a.$ 

Initial profiles of $q$ and $RJ_\phi/B$ as a function of $x = R - R_0$ are shown 
in \rfig{qa1}(b) for $q_a = 2.0, 2.3,  3.0,$ and $3.4$, where $R_0$ is the magnetic axis.
The plasma radius is at $x  = \pm  0.4.$ Horizontal lines are provided to find $q_a,$ and vertical lines are
added to locate $r_s,$ the $q = 2$ rational surface.  As $q_a$ increases, $r_s$ decreases.
The current density is small at $r_s,$ less than $5\%$ of its value on axis. This is stabilizing
for tearing modes. 

The initial profiles in \rfig{qa1}(b) are stable 
according to the  local
criterion \cite{rbm},
\be \Delta = -\pi \sigma \cot(\pi \sigma / 2), \hspace{1.0cm}  \sigma = {J_\phi'}/{B k_\parallel'} \ee
It can be shown that $\sigma \approx  
R J_\phi(r_s) / (2 B),$ assuming $J_\phi' 
 \approx  -J_\phi (r_s) / r_s.$  From  \rfig{qa1}(a), 
$\sigma \approx 0.125.$
In order to have $\Delta > 0,$ it is required that $\sigma > 1.$  
This does not include no wall destabilization.  

Nonlinear runs were carried out with the parameters of MST simulations \cite{mst23}:
Lundquist number $S = 10^5,$ and parallel thermal conductivity $\chi_\parallel = 10 R v_A.$ 
The wall time was taken much shorter than in MST, $S_{wall} = 10^3,$ in order to speed up the
simulations. Here $S_{wall} = \tau_{wall} / \tau_A,$ where  $\tau_{wall}$ is the resistive wall
magnetic penetration time and $\tau_A = R / v_A$ is the \Alf time, with \Alf velocity $v_A$
and major radius $R.$ In MST, $\tau_A = 1.15\times 10^{-6} s.$

Examples of temperature $T$ contours in nonlinear simulations
are shown in \rfig{cont}, 
with $q_a = 3.0.$ 
\rfig{cont}(a) shows $T$ at time $t = 6059 \tau_A,$ with an ideal wall.
The perturbations involve  $(2,1)$ and $(3,2)$  modes. 
\rfig{cont}(b) shows $T$ contours at time
$t = 6390 \tau_A$ with a resistive wall, $S_{wall} = 10^3.$  
 Modes  $(2,1)$ and $(3,2)$  have  larger amplitude
than in the ideal wall case \rfig{cont}(a). There is  a large predominantly $(2,1)$ island structure.
Evidently the initial state was not quite in equilibrium. It relaxed to an unstable  state.
It is clear that the resistive wall has a great effect on the mode evolution.

\begin{figure}[h]
\vspace{.5cm}
\begin{center}
\includegraphics[width=7.5cm]{\figdir/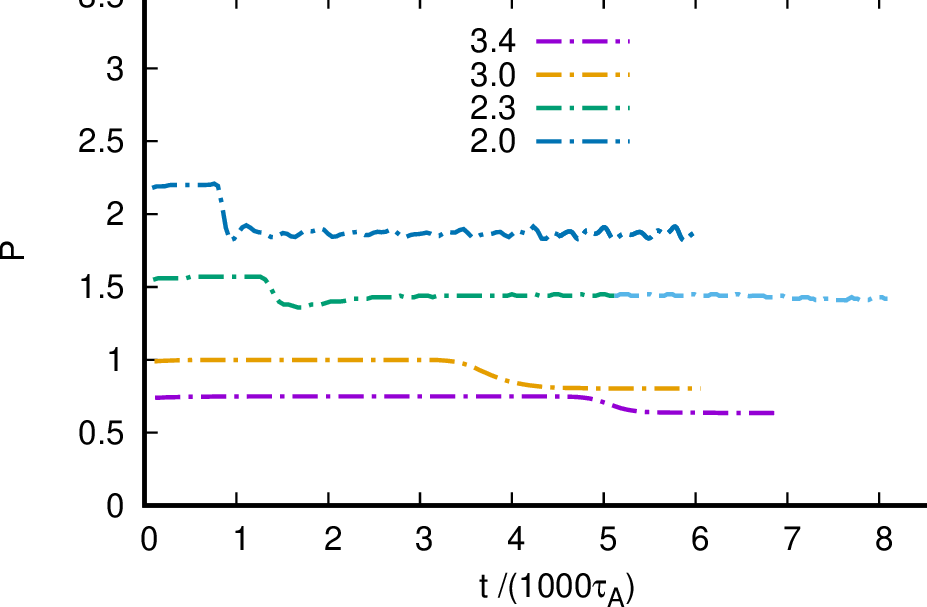}(a)
\includegraphics[width=7.5cm]{\figdir/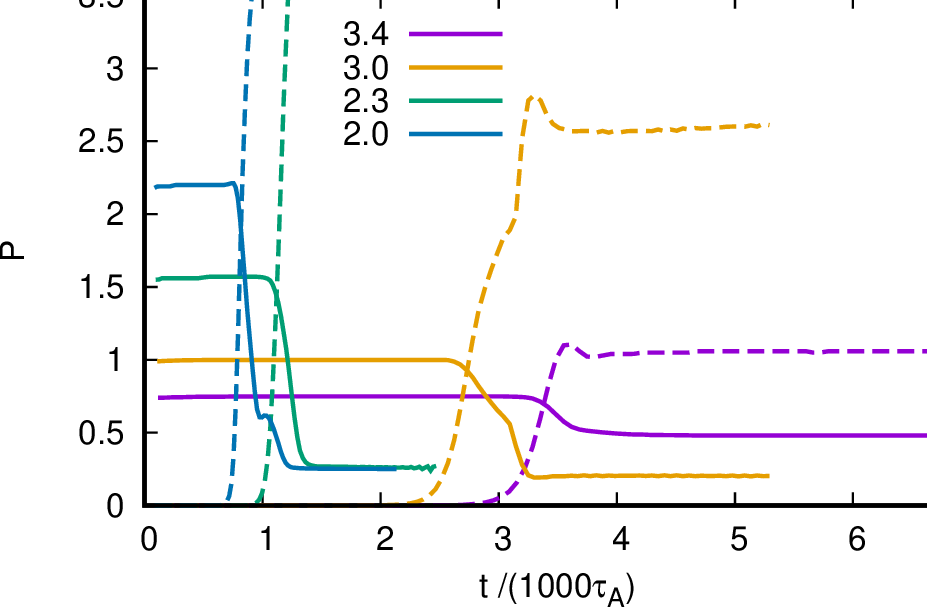}(b)
\end{center}
 \vspace{-.5cm}
\caption{\it
(a) time histories in units of $1000\tau_A$ of total pressure $P$ 
in nonlinear simulations of MST 
with  profiles of \rfig{qa1}(a), with ideal wall. There are only  minor disruptions.
(b) $P$ and $b_n$ as a function of $t / (1000\tau_A)$ for the same initial profiles, for resistive  wall with $S_{wall} = 10^3$.
 }
\label{fig:qa2}
\end{figure}

\rfig{qa2}(a) shows time histories of total pressure $P$ with an ideal wall, 
for cases with $q_a = 3.4, 3.0, 2.3,$
and $2.0.$ 
There are only minor disruptions, with a moderate decrease in $P$. 

With a resistive wall, the results are quite different. 
\rfig{qa2}(b) shows time histories of total pressure $P$ and $10^3 b_n$ for cases with $q_a = 3.4, 3.0,  2.3,$
and $2.0.$ 
Here $b_n$ is the toroidally varying normal magnetic field perturbation at the wall, normalized to
the toroidal magnetic field. With an ideal wall $b_n = 0.$
For $q_a = 3.0, 2.3, 2.0$ there is a  major disruption. This  will be defined 
as a loss of
more than $80 \%$ of the total pressure $P.$
For $q_a = 3.4$ there is a minor disruption.

\begin{figure}[h]
\vspace{.5cm}
\begin{center}
\includegraphics[width=7.5cm]{\figdir/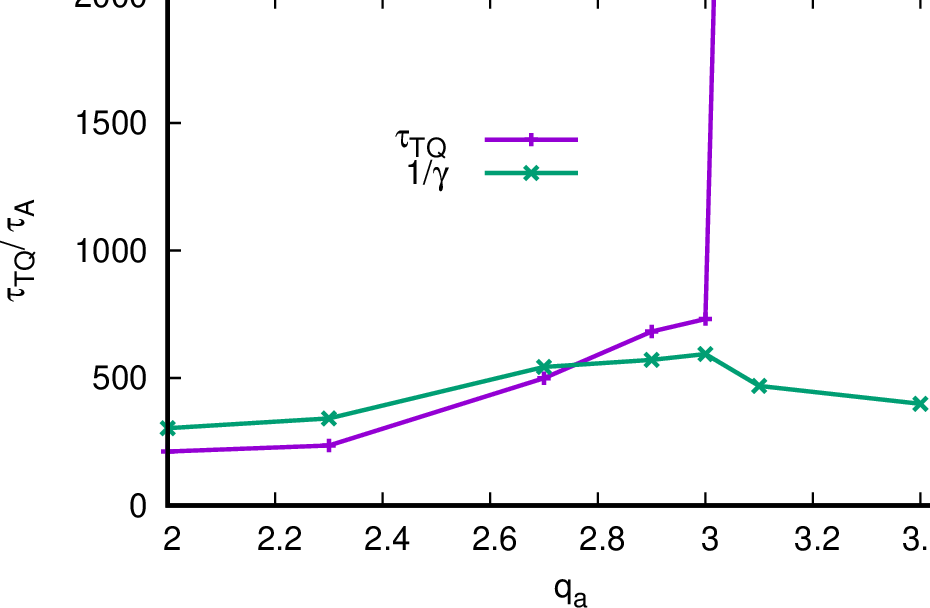}(b)
\includegraphics[width=7.25cm]{\figdir/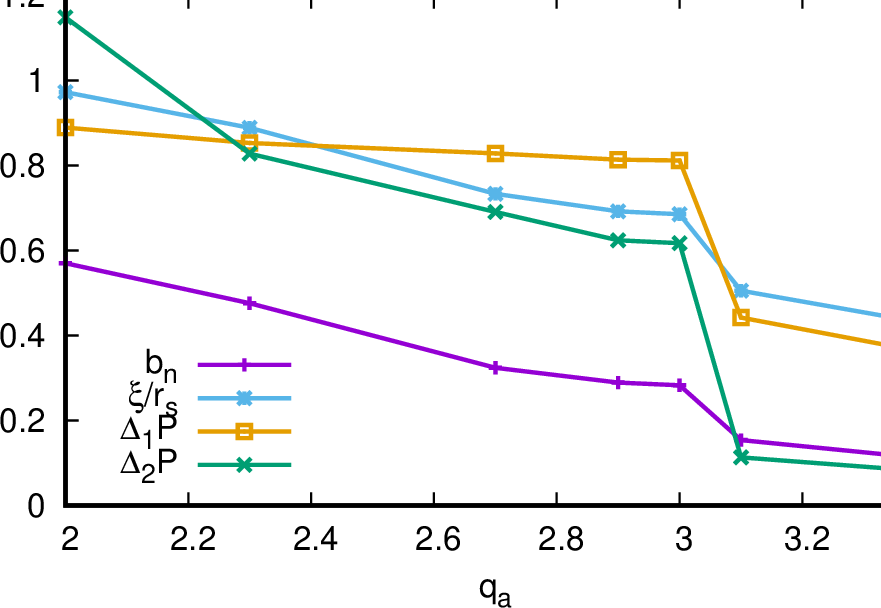}(b)
\end{center}
 \vspace{-.5cm}
\caption{\it
(a) $\tau_{TQ}/\tau_A$ and $1/\gamma$  from time histories, as a function of $q_a.$
There is an abrupt lengthening of the TQ time for $q_a > 3.$ The TQ time  is approximately equal to
the  growth time $1 / \gamma$, which
is typical of RWTMs. Additional time histories with $q_a = 2.7, 2.9, 3.1$ are
included. 
(b) $ 10^2 b_n$, $\Delta_1 P = (1 - P_{min}/P_{max})$, $\Delta_2 P = (P_{max}/P_{min} - 1) / 7$, 
and $\xi = c_0 b_n^{1/2} r_s$ 
as a function of $q_a.$ The quantity $\xi$
is an island width, where $c_0 = 12.9.$ 
These quantities are correlated, with a marked jump at $q_a = 3.$  
}\label{fig:qa3}
\end{figure}

The TQ time $\tau_{TQ}$ varies with $q_a.$ 
For $q_a \ge 3.1$ the
mode is a TM which does not cause a TQ. 
\rfig{qa3}(a) shows $\tau_{TQ}$ and $1 / \gamma $ obtained from \rfig{qa2}(b), as
functions of $q_a.$	
The value of $\tau_{TQ}   = (t_2 - t_9) /(.9 - P_2/P_{max} )$ where $t_2$ is the time in $\tau_A$ units
at which $P = P_2  = .2 P_{max},$ and similarly for $t_9.$  If there is only a minor disruption
for which there is  no value of $t_2,$ then
$t_2$ is replaced by $t_{last},$ and $P_2 = P_{last},$ at the  longest time for that 
particular simulation. This is
an under estimate, but it shows that $\tau_{TQ}$ is much smaller in the RWTM regime
$q_a \le  3.$ At $q_a = 3,$ using the MST \Alf time, $\tau_{TQ} = 0.6 ms.$ This is two orders of
magnitude faster than $\tau_{TQ}$ calculated in the MST experiment \cite{mst23}.
 At higher $q_a$ there should be no RWTMs.
The growth rate $\gamma$ of $b_n$ is calculated.
The relation $\tau_{TQ} = 1/\gamma$ which is well satisfed  for $2 \le q_a \le 3.0$, is
characteristic of RWTMs \cite{jet21}.
The stability boundary is consistent with the linear model \rfig{qa1}(b), for which the onset condition
is $q_a \approx 3.5.$ This is reasonable, considering that the current profiles are
different, and the RWTM regime is small for $q_a > 3.$

\rfig{qa3}(b) shows the maximum value of  $ 10^2 b_n$ and
 $\xi = c_0 b_n^{1/2} r_s, $ 
where $\xi $ is an island width \cite{rutherford}, with
$c_0 = 4 ( 2  R / r_s )^{1/2} (r_w/r_s)^{3/2}.$ The magnetic perturbation at the
rational surface is  $(r_w/r_s)^3 b_n$. Also shown are
two measures of the change in total pressure:
$\Delta_1 P $ $ = 1 - P_{min}/P_{max}$,  and 
$\Delta_2 P $ = $(P_{max}/P_{min} - 1) / 7$,
as a function of $q_a.$ 
There is a marked jump in these
quantities at $q_a = 3.$ The correlation of $\xi$, $\Delta_1 P$, $\Delta_2 P$  for $2 \le q_a \le 3$ suggests 
the transport is advective, with $\delta P \approx  - \xi P'.$ 
The  approximation $\Delta P_2 = \xi/r_s$ has the appropriate amplitude but
a better fit is made using $\Delta P_2 = \xi/r_s.$


\section{Conclusion}

Disruption precursors  have many causes, leading to locked modes in JET and DIII-D.
During precursors, the edge temperature is  reduced, causing the current to  contract.
Disruption onset requires the $q = 2$ rational surface to be sufficiently close to the plasma edge.
This is  consistent with RWTM destabilization. 
Two sets of  model equilibria  are  given which includes  current contraction,
while maintaining constant total current and $q = 1$ on axis.
The first  set is analyzed with linear MHD equations, and solved with ideal wall and no wall boundary
conditions. No wall boundary conditions always make the tearing mode more unstable
than with an ideal wall.
If a tearing mode is stable with an ideal wall and unstable with no wall,
it is a resistive wall tearing mode. 
The model is consistent with experimental disruption  thresholds.
For a sufficiently large  $(2,1)$ rational surface radius $r_s$, 
shrinking the current radius $r_c$ destabilizes the RWTM.
Further shrinking of  $r_c$  stabilizes the RWTM, which exists in a
range of $r_c$ values. 
The model shows that there a maximum value of $q_a$ for which instability is
possible. 

The second, more realistic set of equilibria was used to initialize  nonlinear simulations.
The simulations show a striking difference between ideal and resistive wall.
A sequence of initial states  with different $q_a$ was prepared from an MST equilibrium reconstruction.
These states were contracted inward from the wall, and had very small values of toroidal
current at the $(2,1)$ rational surface. For an ideal wall, minor disruptions occurred. For a
resistive wall, minor disruptions occurred if $q_a > 3.$ For a resistive wall and
edge $q_a \le 3,$ major disruptions occurred. There was a sharp transition in disruptive behavior
at the critical value $q_a = 3.$ 
This indicates that the thermal quench was produced by
RWTMs, and also shows that their onset condition agrees with well known experimental
experience \cite{iter1999}.  

This work provides additional  evidence from theory, simulation, and experimental data that
  disruptions can be caused by resistive wall tearing modes.
MST and ITER have highly conducting walls, so RWTM disruptions are slow.
   RWTMs cause  ``soft disruptions," which  can be passively slowed.
The RWTM disruptions are low $\beta.$ 
High $\beta$ disruptions are resistive wall modes (RWM) \cite{garofalo}, 
which can also be passively slowed.
``Hard disruptions" can occur with  an  ideal wall, and are  caused by making highly unstable equilibria,
using MGI \cite{izzo} , SPI \cite{hu}  or highly unstable initial conditions in simulations \cite{waddell}.

If hard disruptions are avoided, then devices with highly conducting
walls such as ITER \cite{iter21} could experience much milder,
tolerable disruptions than presently predicted \cite{iter1999}.



{\bf Acknowledgement} This work was  supported by USDOE grant DE-SC0020127.

\end{document}